\documentclass[12pt,preprint2]{emulateapj}

\include{epsf}
\usepackage{graphicx}
\usepackage{times}
\newcommand{\hi}{H\,{\sc i}}
\newcommand{\htwo}{{H}$_{2}$}

\newcommand{\msol}{\mbox{${\rm M}_\odot$}}
\newcommand{\hubble}{\mbox{$\rm km\, s^{-1}\, Mpc^{-1}$}}

\newcommand{\mhi}{\mbox{$M_{\rm HI}$}}

\newcommand{\nhi}{\mbox{$N_{\rm HI}$}}
\newcommand{\nhtwo}{\mbox{$N_{\rm H_{2}}$}}

\newcommand{\icmsq}{\mbox{$ \rm cm^{-2}$}}
\newcommand{\fnhi}{\mbox{$f(N_{\rm HI})$}}
\newcommand{\fnhtwo}{\mbox{$f(N_{{\rm H}_2})$}}
\newcommand{\fn}{\mbox{$f(N_{\rm H})$}}
\newcommand{\sfr}{\mbox{$\dot{\rho}_{*}$}}
\newcommand{\sfru}{\mbox{$M_{\odot}\, {\rm yr}^{-1} \, {\rm Mpc}^{-3}$}}
\newcommand{\lya}{Ly$\alpha$}

\slugcomment{{Accepted for publication in ApJ}}

\shorttitle{WHERE IS THE MOLECULAR HYDROGEN IN DLAS?}
\shortauthors{ZWAAN \& PROCHASKA}

\begin{document}

\title{Where is the molecular hydrogen in damped Ly $\alpha$ absorbers?}

\author{Martin A. Zwaan\altaffilmark{1} and Jason X. Prochaska\altaffilmark{2}}

\altaffiltext{1}{European Southern Observatory, Karl-Schwarzschild-Str. 2, 85748 Garching b. M{\"u}nchen, Germany; email: mzwaan@eso.org}
\altaffiltext{2}{UCO/Lick Observatory, University of California, Santa Cruz, 1156 High Street, Santa Cruz, CA 95064; email: xavier@ucolick.org}

\begin{abstract}
We show in this paper why molecular millimeter absorption 
line searches in DLAs have been unsuccessful. We use CO emission line 
maps of local galaxies to derive the \htwo\ column density distribution 
function $f(\nhtwo)$ at $z=0$.  We show that it forms a natural extension 
to \fnhi: the \htwo\ distribution exceeds \fnhi\ at 
$N_{H} \approx 10^{22} \icmsq$ and exhibits
a power law drop-off with slope $\sim -2.5$. 
Approximately 
97\% of the \htwo\ mass density $\rho_{\rm H_2}$ is in systems above 
$\nhtwo=10^{21}\,\icmsq$. 
We derive a value $\rho_{\rm H_2} = 1.1 \times 10^7 h_{70}\,\msol {\rm Mpc}^{-3}$,
which is $\approx 25\%$ the mass density of atomic hydrogen.
Yet, the redshift number density of \htwo\ 
above this \nhtwo\ limit is only $\approx 3\times 10^{-4}$, a factor 150 lower 
than that for \hi\ in DLAs at $z=0$. Furthermore, we show that the median 
impact parameter between a $\nhtwo>10^{21}\,\icmsq$ absorber and the centre 
of the galaxy hosting the \htwo\ gas is only 2.5 kpc. Based on arguments 
related to the Schmidt law, we argue that \htwo\ gas above this column 
density limit is associated with a large fraction of the integral star 
formation rate density. Even allowing for an increased molecular mass 
density at higher redshifts, the derived cross-sections indicate that 
it is very unlikely to identify the bulk of the molecular gas in present 
quasar absorption lines samples. We discuss the prospects for identifying 
this molecular mass in future surveys.
\end{abstract}

\keywords{
galaxies: ISM ---
ISM: molecules ---
ISM: atoms ---
quasars: absorption lines
}

\section{Introduction}

At high redshift, the atomic hydrogen is well surveyed through observations
of damped \lya\ systems (DLAs), quasar absorption line systems
characterized by $\nhi>2 \times 10^{20}\,\icmsq$ 
\citep[e.g.,][]{Prochaska2005}.  These observations
yield the frequency distribution of \hi\ surface density  \fnhi\
and the first moment gives the cosmological mass density of predominantly
neutral, atomic gas.  Several small surveys have been performed to search
for molecular gas associated with DLAs.
Yet, there have been no positive detections of CO and other molecules
in millimeter wavelength absorption line searches \citep[][and references therein]{Curran2004b}. Searches for \htwo\ absorption in DLAs (via the 
Lyman and Werner bands) show a 
success rate of only $\approx 20\%$ and even these sightlines 
have low molecular fractions \citep{Ledoux2003}. 

These low detection rates have drawn into question the relationship between DLAs and star formation at high redshift \citep[e.g.][]{Hopkins2005}. Other galaxy surveys have shown that star formation is very active at these epochs \citep[e.g.,][]{Steidel1999}. Presumably, since stars form in molecular clouds, these star forming galaxies have significant molecular gas mass that has not been discovered through quasar absorption line studies. The obvious conclusion, then, is that these molecular regions have very low covering fraction on the sky and/or contain sufficient columns of dust to obscure any background quasar. 

In this paper we address these issues by
making use of the studies of the CO distribution in nearby galaxies. \citet{Zwaan2005b} recently showed that most DLA properties (luminosities, impact parameters between quasars and DLAs, and metal abundances) are consistent  with them arising in galaxies like those in the local universe. 
Building on this result, we use here information on the molecular content of nearby galaxies, 
to make predictions of the detectability of \htwo\ in DLAs. The \htwo\ distribution in nearby galaxies is generally derived through mapping of the CO(1--0) line at a frequency of 115 GHz. The \htwo\ column densities derived from these observations are typically higher than $5\times 10^{20}\,\icmsq$, which is very close to those reached by millimeter CO absorption line observations in redshifted DLAs against radio-loud background sources.

\section{The H$_{2}$ cross-section}

The largest sample of high spatial resolution CO(1--0) maps of nearby galaxies is the BIMA SONG sample presented by \citet{Helfer2003}. The sample consists of 44 optically selected galaxies with Hubble types Sab to Sd at a median distance of 11.9 Mpc, and is selected without reference to CO or infrared brightness. {The fact that this sample does not include early types and irregulars is not expected to introduce an important bias in our calculations. CO has been detected in a number of elliptical galaxies \citep{Young2002}, but their contribution to the total \htwo\ mass density is expected to be minimal. Irregulars contribute approximately 
20\% to the total \hi\ mass density \citep{Zwaan2003}, but their  ratio of \htwo\ mass to \hi\ mass is down by almost a factor ten compared to the average over all galaxy types
\citep{Young1989}, which implies that their contribution to the \htwo\ mass density 
is also marginal. We refer to \citet{Zwaan2005b} for a discussion on why optically selected galaxies are a fair sample 
to compare  with absorption line statistics. }
 
We used the combined interferometric
and single dish maps to calculate \htwo\ column density maps, using a CO/\htwo\ conversion factor of $2.8 \times 10^{20} \,\rm cm^{-2} {(K \,km\, s^{-1})}^{-1}$ {\citep{Kennicutt1998}.} \citet{Helfer2003} report that the lowest significant isolated \htwo\ densities at $3\sigma$ in these maps are approximately $13.7 \,\msol \rm \, pc^{-2}$, which corresponds to an \nhtwo\ limit of $8.5 \times 10^{20} \rm cm^{-2}$. 
The typical spatial resolution of the $6''$ beam is 350 pc at the median distance of the sample.

We calculate the \htwo\ column density distribution $f(\nhtwo)$ using a method 
analogous to that for \hi\ set out by \citet{Zwaan2005b}:
 \begin{equation}
f(\nhtwo)=\frac{c}{H_0} \frac{\sum_i  \Phi({M_{B}}_i) w({M_{B}}_i) A_i (\log \nhtwo)}{\nhtwo \, \ln 10 \,\Delta \log \nhtwo  }.
\end{equation}
 Here, $\Phi({M_{B}}_i)$ is the space density of galaxy $i$
measured through the optical luminosity function as measured by 
\citet{Norberg2002b}, {with Schechter parameters:
$M^{*}_{B}- 5 \log h_{70}=-20.43$, $\alpha=-1.21$,  and $\Phi^{*}= 5.5 \times 10^{-3} h_{70}^{3} {\rm Mpc}^{-3}$.}
The function $w({M_{B}}_i)$ is a weighting function that takes into account the varying number of galaxies across the full stretch of $M_{B}$,
and is calculated by taking the reciprocal of the number of galaxies in the range ${M_{B}}_i-\Delta/2$ to  
${M_{B}}_i+\Delta/2$, where $\Delta$ is taken to be 0.25. 
 $A_i(\log \nhtwo$) is the area function that describes for
galaxy $i$ the area in $\rm Mpc^{2}$
corresponding to a column density in the range $\log \nhtwo$ to $\log \nhtwo+\Delta \log \nhtwo$.
{In practice, this is simply calculated by summing for each galaxy 
the number of pixels in a certain $\log \nhtwo$ range multiplied by the physical
area of a pixel.}
Finally, $c/H_0$ converts the number of systems per Mpc to
that per unit redshift.

The BIMA SONG galaxies are selected to be less inclined than 70$^\circ$. In order to achieve a $f(N)$ measurement for randomly oriented galaxies, we de-projected all galaxies to face-on assuming that the \htwo\ gas is optically thin, and subsequently recalculated the column density distribution function for ten inclinations $i$ evenly spaced in $\cos(i)$ between 0 and 1.
{The final \fn\ was taken to be the average of these ten distribution functions. This procedure only makes a small modification to the \fn\ calculated from the \htwo\ distributions uncorrected for inclination effects.}

\begin{figure}
\includegraphics[width=8.5cm,trim=0cm 2cm 0cm 0cm]{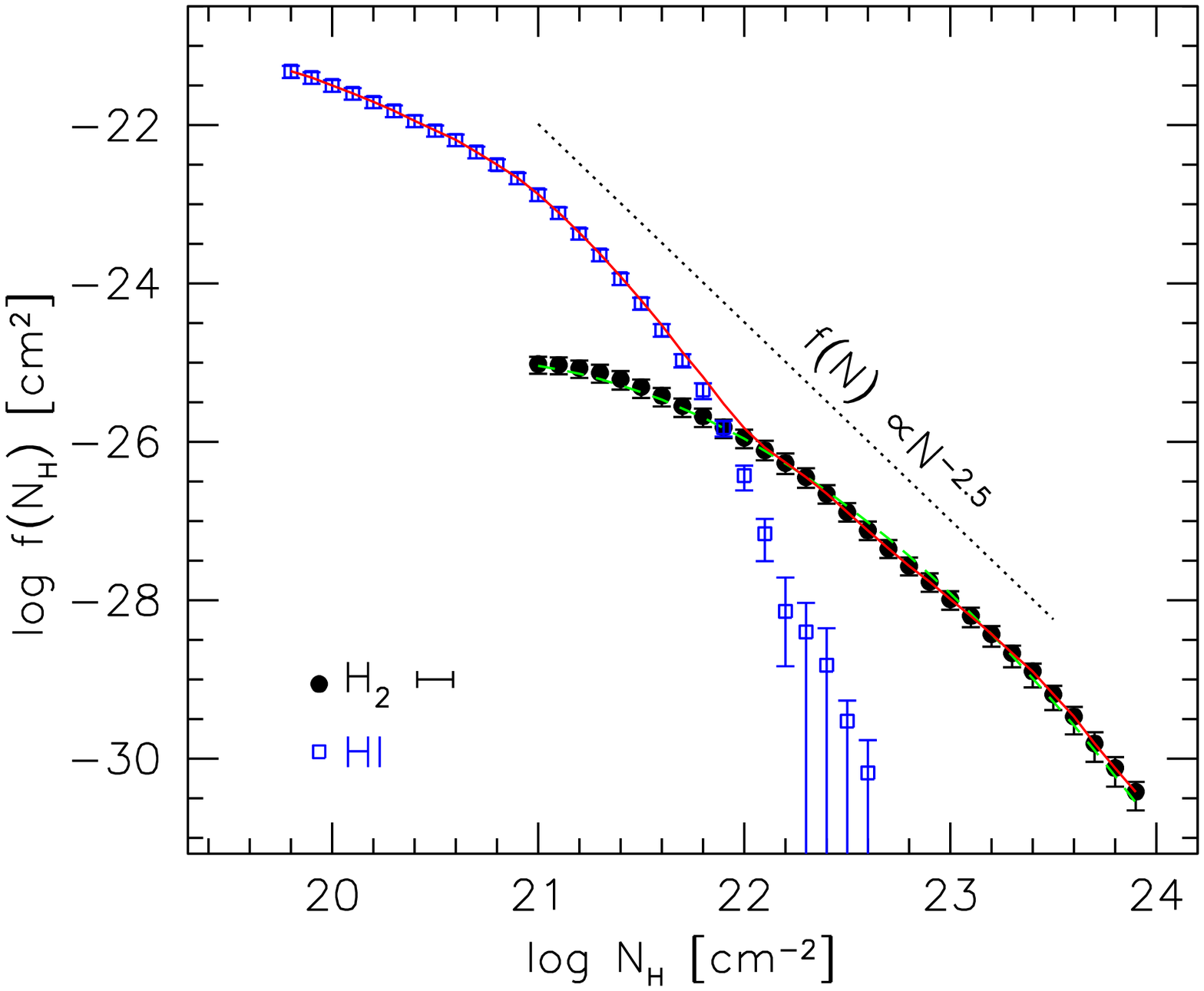}
\caption{The column density distribution function of \hi\ and \htwo\ at $z=0$. The \hi\ curve is adopted from \citet{Zwaan2005b}, the \htwo\ curve is measured from the CO emission line maps from the BIMA SONG sample. Column densities are expressed in atoms per cm$^{2}$, also for \htwo. The solid line is the summed \fn. 
The horizontal errorbar indicates the uncertainty in \fnhtwo\ if the CO/\nhtwo\ conversion factor is uncertain by 50\%. 
The \fnhtwo\ can be fitted very well with a log-normal distribution, where $\mu=20.6$, $\sigma=0.65$ and the normalization is $1.1\times 10^{-25}\, {\rm cm}^{2}$, as indicated 
by the dashed line.
\label{fnh2.fig}}
\end{figure}

Figure \ref{fnh2.fig} shows the resulting column density distribution function \fnhtwo, together with \fnhi\ from \citet{Zwaan2005b}. We note that 
the horizontal axis represents the atom surface density, which in the case of \htwo\ is equal to  $2\nhtwo$. {The errorbars are 1$\sigma$ uncertainties and include counting statistics and the uncertainty in the optical luminosity function. The uncertainty in the CO/\htwo\ conversion factor could introduce the largest error in our \fnhtwo. The horizontal errorbar
indicates the uncertainty in \fnhtwo\ if this conversion factor is uncertain by 50\%.}
{The \fnhtwo\ can be fitted very well with a log-normal distribution: $f(N)=$
$f^{*}\exp{-[(\log N-\sigma)/\mu]^2/2}$
, where $\mu=20.6$, $\sigma=0.65$ and the normalization $f^*$ is $1.1\times 10^{-25}\, {\rm cm}^{2}$.}
The distribution function of \htwo\ column densities seems to follow a natural extension of the \hi\ distribution function, in such a way that the summed \fn\ roughly follows a power-law distribution $N_{\rm H}^{-2.5}$
between $\log N_{\rm H}=21$ and $\log N_{\rm H}=24$. 
The two distribution functions cross at $\log N_{\rm H}\approx 22$, 
which is the approximate column density associated with 
the conversion from \hi\ to \htwo\ \citep[e.g.][]{Schaye2001b}.

By taking the integral over \fnhtwo\ multiplied by \nhtwo, we find the total \htwo\ mass density at $z=0$ to be $\rho_{{\rm H}_2}=1.1 \times 10^{7} \msol \, {\rm Mpc}^{-3}$, which is approximately one 
quarter of $\rho_{\rm HI}(z=0)$ \citep{Zwaan2005a}.
\cite{Keres2003} found  $\rho_{{\rm H}_2}=(2.0 \pm 0.7) \times 10^{7} \msol \, {\rm Mpc}^{-3}$, where we converted their value to $H_0=70 \,\hubble$ and used the same CO/\htwo\ conversion factor as we used. 
 {This difference might be partly due to the fact that the BIMA SONG sample does not include many dwarf galaxies, although judging from the \citet{Keres2003}
CO mass function, these galaxies contribute only very little to $\rho_{{\rm H}_2}$.
Another reason could be that the} BIMA SONG sample is optically selected, whereas Keres' sample \citep{Young1989} predominantly consists of FIR-selected galaxies, which can cause a bias toward CO-rich galaxies \citep[see e.g.,][]{Solomon1988}. \citet{Casoli1998} used a larger sample than \citet{Young1989}, and took into account CO non-detections, to find much lower values of $M_{{\rm H}_2}/\mhi$, which are fully consistent with our derived value of $\rho_{{\rm H}_2}/\rho_{\rm HI}$.

\begin{figure}
\includegraphics[width=8.5cm,trim=0.4cm 4cm 4cm 0cm]{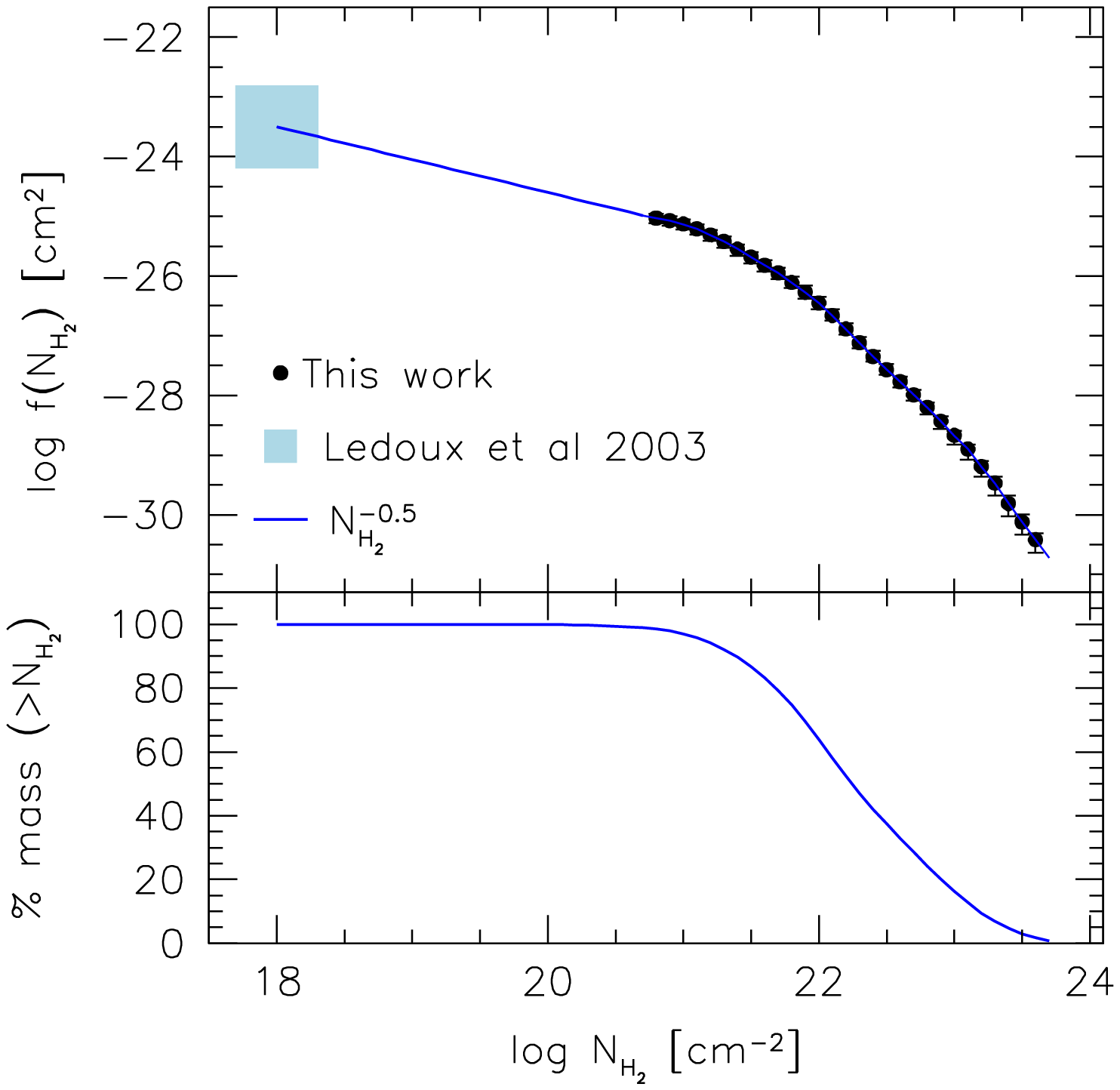}
\caption{{\em Top:\/} The column density distribution function of \htwo, where the circles show \fnhtwo\ at $z=0$ derived from the BIMA SONG CO maps, and the grey square indicates the  value at much lower column densities, estimated from the \htwo\ absorption data from \citet{Ledoux2003}. These latter data are from systems at redshifts $z\approx 2$, which implies that the derived \fnhtwo\ point is probably an upper limit.
{\em Bottom:\/} The cumulative \htwo\ mass distribution in column densities $>\nhtwo$. Here, we have  includes the $\nhtwo^{-0.5}$ extension as shown in the top panel. Approximately 97\% of the \htwo\ molecules are in column densities $\nhtwo>10^{21}\,\icmsq$. 
\label{fnledoux.fig}}
\end{figure}

What fraction of the cosmic \htwo\ mass density do we miss below the \nhtwo\ detection limit of $8.5 \times 10^{20} \rm cm^{-2}$? 
The \fnhtwo\ appears to flatten off at the lowest column densities, which implies that the contribution of low \nhtwo\ is low, unless the \fnhtwo\ rises steeply below our detection limit.
To test this, we make use of the \htwo\ absorption line survey 
in DLAs by \citet{Ledoux2003}. These authors report a detection 
rate of 13 to 20\%, with \htwo\ column densities typically in the range $\log \nhtwo=17$ to $18.5$. 
Based on their statistics and the measured \fnhi\ of DLAs at $z=0$ from \citet{Zwaan2005b}, we estimate that $\log f(\nhtwo=10^{18} \icmsq)=-23.5$ at $z=0$, conservatively assuming that the detection statistics of \htwo\ absorption has not evolved since $z\approx 2$. 
Extrapolating our measured \fnhtwo\
to this value, we find that $\fnhtwo\propto\nhtwo^{-0.5}$
(see Figure~\ref{fnledoux.fig}). 
Now, by integrating $\int\nhtwo\fnhtwo d\nhtwo$ we find the total \htwo\ mass as a function of \nhtwo. 
From this we derive that the mass contained in systems $\log\nhtwo<21$ is only 3\% of the total \htwo\ mass density. 
This implies that our results presented in this paper apply to 
roughly 97\% of the total number of  \htwo\ molecules in the universe.

Beam smearing might lead to an overestimation of the cross-sections 
at low $\nhtwo$ and an underestimte of $\fnhtwo$ at large $\nhtwo$.
Because a main result of the next section
is that the \htwo\ cross-sections are small, we have ignored these effects
for now.

\section{Where to find the molecules}
The redshift number density $dN/dz$ of \htwo\ above the \nhtwo\ limit
of $10^{21}~\icmsq$ can be calculated from \fnhtwo\ by summing over all column densities larger than this limit. We find that $dN/dz=3 \times 10^{-4}$, which is approximately 150 times lower than the corresponding value for \hi\ above the DLA limit \citep{Zwaan2005b}. Taking this result at face value, this would imply that at $z=0$ only one in every $\sim$ 150 DLAs is expected to show strong CO absorption lines.

\begin{figure}
\includegraphics[width=8.5cm,trim=0cm 4cm 0cm 0cm]{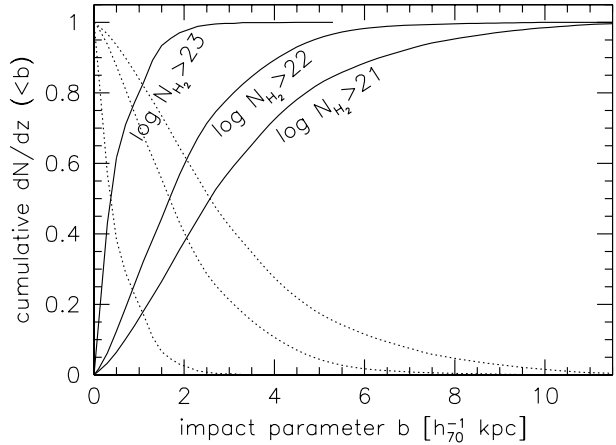}
\caption{The normalized $z=0$ cumulative distribution function of impact parameters of \htwo\ cross-section selected systems above different \htwo\ column density limits. The solid is the distribution for impact parameters $<b$, the dotted line for $>b$.
\label{impact.fig}}
\end{figure}

In Figure~\ref{impact.fig} we show how close to the centre of galaxies high column density \htwo\ is to be found. This figure shows the normalized cumulative distribution function of impact parameters between the centers of galaxies and the positions where the \htwo\ absorption is measured. 
We constructed this diagram from the BIMA SONG CO maps, and using a weighting scheme similar as discussed for the evaluation of \fnhtwo. 
Apparently, the median impact parameter at which 
an $\nhtwo>10^{21}\,\icmsq$ absorber is expected is only 2.5 kpc and 90\% of the cross-section is at impact parameters smaller than 6.5 kpc. For higher \htwo\ column densities, these impact parameters are even smaller. Applying these findings to the DLA population, millimeter CO observers need to select those DLAs that arise within 2.5 kpc of the centers of galaxies to have a 50\% probability of identifying an absorption line corresponding to $\nhtwo>10^{21}\,\icmsq$. None of the {\em identified} DLA host galaxies have such small impact parameters, except perhaps two cases studied by \citet{Rao2003} for which only upper limits to the impact parameters could be given due to blending with the background quasar light.

Our conclusions are qualitatively supported by observations of redshifted molecular lines. To date, only four redshifted molecular absorbers have been identified \citep[][and references therein]{Wiklind1999}, two of which arise in gravitationally lensed intervening galaxies and two originate in the galaxy that also hosts the radio source against which the absorption is seen. In all cases the sightline through the high \htwo\ column passes through the galaxies at small impact parameters.

These cross-section arguments already explain the very low detection rate of molecules in DLAs, but there are  additional effects that hinder the detection of molecules.
Molecular hydrogen forms on the surface of dust grains. The regions in galaxies containing most of the universe's \htwo\ molecules are therefore likely to be dusty, causing a higher optical extinction of the background sources, which, in turn, might lead to them dropping out of magnitude-limited surveys. 
Obviously, radio-selected quasars do not suffer from this bias and are potentially good candidates against which to find high \nhtwo\ absorbers. However, $dN/dz(\nhtwo)$ is so low that identifying such an absorber is highly unlikely in current radio-selected quasars samples. For example, the redshift interval covered by the CORALS survey \citep{Ellison2001} of 
radio-loud quasars is only $\Delta z=55$ for \lya, 
corresponding to $\Delta z \approx 100$ for \htwo. 
The redshift interval required for identifying a high \nhtwo\ 
absorber should be 
on the order of $(dN/dz)^{-1}=3300$ at $z=0$, or approximately 600 at $z=2-3$, taking into account the cosmological variation of the absorption distance interval $dX$. Using this method to obtain sufficient statistics to measure the molecular mass density at intermediate and high redshifts, would require radio source samples larger than currently available. 

Finally, we address the question whether there might exist a significant amount of \htwo\ gas not associated with DLAs. CO mapping of nearby galaxies has shown that many galaxies show a depression in the \hi\ distribution where the \htwo\ column densities are highest \citep[see e.g.][]{Wong2002}.  
Presumably, in regions where the molecular densities are highest, most of the \hi\ has been converted to \htwo. We use the \hi\ and CO maps of seven nearby galaxies studied by \citet{Wong2002} to test how much \htwo\ exists below the \hi\ DLA limit. We compare the \hi\ and CO maps smoothed to the same spatial resolution of $13''$ to $23''$, and find that the mass fraction of \htwo\ in regions where $\nhi<2\times 10^{20}\,\icmsq$ ranges from 0 to 40\%. Averaged over all pixels from the total sample, this fraction is 6\%. In many cases the fraction is underestimated because the highest  \nhtwo\ is smoothed over larger regions. Of course, this small sample may not be representative of the total galaxy population at $z=0$, but at least this exercise shows that a small fraction of the cosmic \htwo\ density must be found in sub-DLAs. 
One might identify such absorbers by searching for a sub-DLAs with
abnormally high metallicities, i.e.\ this could indicate
absorbers where the hydrogen is not predominantly atomic. An example of a high metallicity sub-DLA has recently been found by \citet{Peroux2006}.

\section{Implications for the star formation rate density}
\citet{Lanzetta2002} and \citet{Hopkins2005} recently estimated the star formation rate density (SFRD) in DLA systems by applying the `Schmidt law' of star formation to the \hi\ column density distribution function \fnhi. The Schmidt law is defined in local galaxies and states
that the star formation rate correlates very well with total neutral gas surface density $\Sigma_{\rm HI}+\Sigma_{\rm H_{2}}$ to the power 1.4, as was demonstrated by \citet{Kennicutt1998}. 
Here, we wish to investigate whether the SFRD measurements based on \hi\ alone give
meaningful results, or whether \htwo\ should be taken into account for a reliable 
SFRD measurement.  Since presently we only know \fnhtwo\ at $z=0$, we 
cannot improve the measurements of \citet{Hopkins2005} at high redshift by including \htwo. Our aim is simply to test the validity of the method and discuss the implications of including molecules.

We start with estimating what fraction of the SFRD is actually contributed by those regions where the \htwo\ column is higher than the \hi\ column. To this end, we make use of Equation 4 from \citet{Hopkins2005} that relates the SFRD to the \fn, and apply it to our measurement of $f(\nhtwo)$, and \fnhi\ from \citet{Zwaan2005b}. We make one important modification in that we convert our $z=0$ gas densities to those that would be observed if the gas disks were observed `face-on', assuming that the \hi\ and \htwo\ layers are optically thin. At $z=0$ the \fn\ is a result of a randomly oriented population of galaxies and it is easy to see that the highest column densities are mostly the result of highly inclined disks. 
However, the Schmidt law is valid for face-on gas densities. De-projecting the disk implies that the surface area increases but the column densities decrease. The net result is that that the \fn\ as it is normally defined will overestimate the SFRD. At $z=0$, we find that the SFRD is overestimated by $\sim 40$\% if the regular \fn\ is used, instead of the de-projected \fn\ for face-on disks. The de-projected \fn\ follows a nearly exponential behaviour between $\log N_{\rm H}=20.5$ and  $\log N_{\rm H}=23.5$, and can be fitted with the simple equation $f(N_{\rm H})=2\times 10^{29} N_{\rm H}^{-2.5}$.

\begin{figure}
\includegraphics[width=8.5cm,trim=0cm 4cm 0cm 0cm]{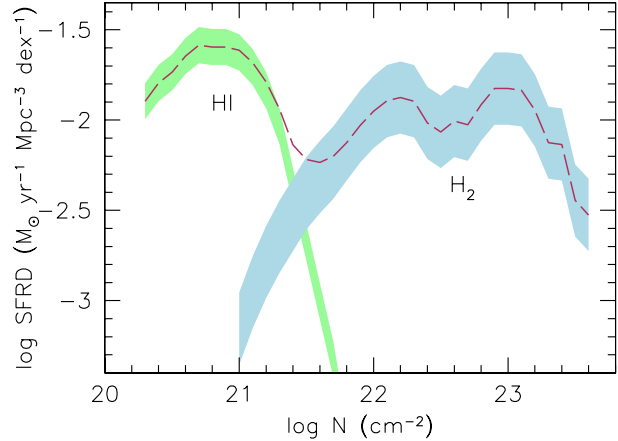}
\caption{The implied star formation rate density as a function of face-on \hi\ and \htwo\ column density as derived from the \citet{Kennicutt1998} star formation law. Grey areas indicate approximate uncertainties.
\label{sfrd_N.fig}}
\end{figure}

Our results are presented in Fig.~\ref{sfrd_N.fig}, which shows the implied SFRD as a function of \hi\ and \htwo\ face-on column density. We see that the \hi\ and \htwo\ column densities contribute approximately equally to the total SFRD. 
The other conclusion from this exercise is that the total derived SFRD at $z=0$
is much higher than that derived from H$\alpha$ and [O{\sc ii}] measurements 
[$\sfr(z=0)\approx 0.02 \sfru$, \citealt{Hopkins2004}].
This contradicts the findings of \citet{Hopkins2005} who conclude that at $z=0$ the SFRD \sfr\ derived from the Schmidt law and the measured \fnhi\ agrees well with the direct measurements. The origin of this contradiction is that \citet{Hopkins2005} used the \fnhi\ measurements of \citet{Ryan-Weber2003}, which were later corrected upwards with a factor 3. If we use the corrected values \citep{Ryan-Weber2005a} or the measurement from \citet{Zwaan2005b}, we find that $\sfr(z=0)=0.035 \sfru$, a factor two higher than the median of the directly measured values at $z=0$. Using the more realistic face-on \fn, and including both \hi\ and \htwo\ in the analysis, we find that $\sfr(z=0)=0.044\sfru$,
{also higher than the nominal value at $z=0$.} 
Based on the  face-on \fnhtwo\ only, we find $\sfr(z=0)=0.022 \sfru$, in good agreement with the direct measurements \citep[see also][]{Wong2002}.

Why is the value of \sfr\ at $z=0$ overestimated by a factor $2-3$ when derived using the \fn\ and the Schmidt law? The answer lies in the fact that the Schmidt law is defined for the `star-forming disks', and not for the regions of the \hi\ layer outside this area.  It was shown by \citet{Kennicutt1989} that star formation only occurs when the gas density exceeds the critical threshold density, which depends on the velocity dispersion of the gas and the galaxy's rotation curve shape and amplitude. From the \fn\ alone it is impossible to determine what the critical threshold gas density is, because this density varies between galaxies and within galaxies. Therefore, applying the Schmidt law to all regions in the local universe where the \nhi\ exceeds $\sim 10^{20}\,\icmsq$ will grossly overestimate \sfr. 
{[As an extreme example, consider NGC 2915,  \citep{Meurer1996} where the HI disk  is many times larger than the optical disk.]}

Furthermore, within galaxies the areas with the highest SFRs are often mostly molecular,
and in some cases \nhi\ declines in regions of high SFR \citep{Martin2001,Rownd1999,Wong2002}. 
In regions where $\log \nhtwo>21$ the critical density is typically exceeded. Consequently, for the purpose of estimating \sfr, the \fnhtwo\ is probably a more reliable estimator, whereas \fnhi\ only gives an upper limit to that fraction of \sfr\ contributed by regions that are mostly atomic. 

{In our analysis we treat the \hi\ and \htwo\ independently. Ideally,
we would use \hi\ and \htwo\ measurements from the same large galaxy
sample, but unfortunately such a sample is not available. Our analysis
probably slightly {\em under}estimates the total SFRD: in regions
where the \hi\ and \htwo\ columns are equal, the SFR would be
$2^{1.4}/2=1.3$ times higher if it were derived from the summed gas
density instead of from the individual densities. In most regions,
where the two column densities are different, the SFR is
underestimated by a much smaller fraction if the \hi\ and \htwo\ are
treated independently.}

In summary, we have shown here that \fnhi\ cannot be used to derive a meaningful SFRD at $z=0$ and therefore probably also not at higher redshifts. Only in the unlikely situation that DLA galaxies were truly molecule free, the \fnhi\ method would give a meaningful upper limit to the SFRD contributed by DLAs. Interestingly, the {\em shape} of \fnhi\ has not evolved\footnote{{Note however that \citet{Rao2005} find a much flatter \fnhi\ distribution at $0.1<z<1.7$, based on Mg{\sc ii}-selected systems.}} between $z=4$ and $z=0$ \citep[cf.][]{Zwaan2005b,Prochaska2005}, only the {\em normalization} (i.e., $\Omega_{\rm HI}$) has decreased with approximately a factor two. Based on the \citet{Hopkins2005} parameterization, this implies that \sfr\ {as measured through DLAs} also has dropped only approximately a factor two over this redshift range. {This is clearly at odds with the observations, unless one is willing to assume that the systems holding the bulk of the neutral 
gas account for only a small fraction of the cosmic SFRD}. Furthermore, even at $z=0$ the SFRD is dominated by regions where  \nhtwo\ exceeds \nhi\ {\citep[see also][]{Wong2002}}. Given the observational fact that \sfr\ evolves much faster than $\Omega_{\rm HI}$, this implies that the cosmic \htwo\ mass density was much higher in the past, or the laws of star formation were different.

{In any case, the contention of \citet{Hopkins2005} that DLAs cannot make up for the observed SFRD at high redshift should be reviewed. Clearly, to evaluate the SFRD contributed
by DLAs one should take into account their molecular content, which we have shown in the previous paragraph is extremely difficult to determine.
In the absence of a measurement of \fnhtwo\ at higher redshifts, 
more direct approaches such as the C{\sc ii}* method \citep{Wolfe2003} 
or models of the star formation history \citep{Dessauges2003} are probably more
useful in constraining \sfr\ from DLAs. }

\section{Conclusions}
We have used observations of CO in nearby galaxies to make predictions on the detection rate of \htwo\ in DLAs. We derived a column density distribution function \fnhtwo\ and find that it forms a natural extension of \fnhi\ at higher hydrogen column densities. The inferred redshift number density $dN/dz$ of $\nhtwo>10^{21}\,\icmsq$ is $3 \times 10^{-4}$, which is a factor 150 lower than the corresponding number for DLAs. This implies that of the total number of DLAs currently known, only a handful are expected to show high column density molecular absorption lines. This absorption is expected to arise at impact parameters smaller than 6.5 kpc in 90\% of the cases. Approximately 97\% of the cosmic \htwo\ mass density resides in this dense, central gas, which also is associated with the bulk of the cosmic star formation rate density \sfr. The detectability of this molecular gas is further reduced by the facts that  {\em i)\/} the high column density \htwo\ is likely associated with high dust columns, which obscures background quasars and could take them out of magnitude-limited surveys, and {\em ii)\/} a small fraction ($\la 10$\%) of the \htwo\ is associated with sub-DLA \lya\ absorption.  
 
{We apply the Schmidt law for star formation to \fnhtwo\ and \fnhi\ and find
that molecules should be taken into account to derive meaningful values
for the cosmic star formation rate density \sfr. Even though the cross section of high column density \htwo\ is low,  this gas is associated with a large fraction of \sfr. Since the molecular gas is difficult to detect in absorption,
the method of using the Schmidt law and \fn\ to derive \sfr\ remains of limited use.}

The cold atomic gas content of the universe at redshifts $z\sim 1-5$ has successfully been determined from blind surveys for DLA absorbers. What are the prospects for measuring 
the molecular mass density at intermediate and high redshifts using similar techniques?
With present technology blind molecular absorption line surveys against radio-loud background sources seem difficult given the required redshift interval. The prospects for future instruments seem better. For example, with the Atacama Large Millimeter Array\footnote{http://www.alma.info/} (ALMA) the CO(3--2) line 
corresponding to $\log\nhtwo>21$ should be detectable at $z=2$ to $3$ in one minute against a 50 mJy background source. 
 A survey of
several days should be able to turn up a useful number of high \nhtwo\ absorbers.
{Such a survey would by simultaneously sensitive to HCO$^+$ (a good \htwo\ tracer) at lower redshifts.}
The Square Kilometer Array\footnote{http://www.skatelescope.org/} (SKA) should be able to detect the CO(1--0) and
HCO$^+$(1--0) lines at redshifts $z>3.6$ and $>2.6$, respectively. 
The expected noise levels are much lower than for ALMA, implying that this instrument will be ideal for high-$z$ molecular absorption line surveys, giving good statistics for measuring $\Omega_{{\rm H}_{2}}$.

\acknowledgments
We wish to thank C. P{\'e}roux, H.-W. Chen, A. Wolfe, L. Blitz, J. Liske, N. Bouch{\'e}, and S. Rao for helpful criticism
on the manuscript and T. Wong for making available 
\hi\ and CO maps in electronic format. 
We also thank the anonymous referee for a quick and very detailed report.
JXP acknowledges support
from NSF grants AST 03-07408 and AST 03-07824.  
This work was fostered by ESO's Scientific Visitor Programme.


\begin{thebibliography}{33}
\expandafter\ifx\csname natexlab\endcsname\relax\def\natexlab#1{#1}\fi

\bibitem[{{Casoli} {et~al.}(1998){Casoli}, {Sauty}, {Gerin}, {Boselli},
  {Fouque}, {Braine}, {Gavazzi}, {Lequeux}, \& {Dickey}}]{Casoli1998}
{Casoli}, F. {et~al.} 1998, \aap, 331, 451

\bibitem[{Curran {et~al.}(2004)Curran, Murphy, Pihlstr{\"o}m, Webb, Bolatto, \&
  Bower}]{Curran2004b}
Curran, S.~J., Murphy, M.~T., Pihlstr{\"o}m, Y.~M., Webb, J.~K., Bolatto,
  A.~D., \& Bower, G.~C. 2004, \mnras, 352, 563

\bibitem[{{Dessauges-Zavadsky} {et~al.}(2003){Dessauges-Zavadsky}, {D'Odorico},
  {Prochaska}, {Calura}, \& {Matteucci}}]{Dessauges2003}
{Dessauges-Zavadsky}, M., {D'Odorico}, S., {Prochaska}, J.~X., {Calura}, F., \&
  {Matteucci}, F. 2003, Elemental Abundances in Old Stars and Damped
  Lyman-{$\alpha$} Systems, 25th meeting of the IAU, JD 15, 15

\bibitem[{Ellison {et~al.}(2001)Ellison, Yan, Hook, Pettini, Wall, \&
  Shaver}]{Ellison2001}
Ellison, S.~L., Yan, L., Hook, I.~M., Pettini, M., Wall, J.~V., \& Shaver, P.
  2001, \aap, 379, 393

\bibitem[{Helfer {et~al.}(2003)Helfer, Thornley, Regan, Wong, Sheth, Vogel,
  Blitz, \& Bock}]{Helfer2003}
Helfer, T.~T., Thornley, M.~D., Regan, M.~W., Wong, T., Sheth, K., Vogel,
  S.~N., Blitz, L., \& Bock, D.~C.-J. 2003, \apjs, 145, 259

\bibitem[{Hopkins(2004)}]{Hopkins2004}
Hopkins, A.~M. 2004, \apj, 615, 209

\bibitem[{Hopkins {et~al.}(2005)Hopkins, Rao, \& Turnshek}]{Hopkins2005}
Hopkins, A.~M., Rao, S.~M., \& Turnshek, D.~A. 2005, \apj, 630, 108

\bibitem[{{Kennicutt}(1989)}]{Kennicutt1989}
{Kennicutt}, R.~C. 1989, \apj, 344, 685

\bibitem[{Kennicutt(1998)}]{Kennicutt1998}
Kennicutt, R.~C. 1998, \apj, 498, 541

\bibitem[{Keres {et~al.}(2003)Keres, Yun, \& Young}]{Keres2003}
Keres, D., Yun, M.~S., \& Young, J.~S. 2003, \apj, 582, 659

\bibitem[{Lanzetta {et~al.}(2002)Lanzetta, Yahata, Pascarelle, Chen, \&
  Fern{\'a}ndez-Soto}]{Lanzetta2002}
Lanzetta, K.~M., Yahata, N., Pascarelle, S., Chen, H.-W., \&
  Fern{\'a}ndez-Soto, A. 2002, \apj, 570, 492

\bibitem[{Ledoux {et~al.}(2003)Ledoux, Petitjean, \& Srianand}]{Ledoux2003}
Ledoux, C., Petitjean, P., \& Srianand, R. 2003, \mnras, 346, 209

\bibitem[{{Martin} \& {Kennicutt}(2001)}]{Martin2001}
{Martin}, C.~L., \& {Kennicutt}, R.~C. 2001, \apj, 555, 301

\bibitem[{{Meurer} {et~al.}(1996){Meurer}, {Carignan}, {Beaulieu}, \&
  {Freeman}}]{Meurer1996}
{Meurer}, G.~R., {Carignan}, C., {Beaulieu}, S.~F., \& {Freeman}, K.~C. 1996,
  \aj, 111, 1551

\bibitem[{{Norberg} {et~al.}(2002)}]{Norberg2002b}
{Norberg}, P., {et~al.} 2002, \mnras, 336, 907

\bibitem[{{P{\'e}roux} {et~al.}(2006){P{\'e}roux}, {Kulkarni}, {Meiring},
  {Ferlet}, {Khare}, {Lauroesch}, {Vladilo}, \& {York}}]{Peroux2006}
{P{\'e}roux}, C., {Kulkarni}, V.~P., {Meiring}, J., {Ferlet}, R., {Khare}, P.,
  {Lauroesch}, J.~T., {Vladilo}, G., \& {York}, D.~G. 2006, astro-ph/0601079

\bibitem[{{Prochaska} {et~al.}(2005){Prochaska}, {Herbert-Fort}, \&
  {Wolfe}}]{Prochaska2005}
{Prochaska}, J.~X., {Herbert-Fort}, S., \& {Wolfe}, A.~M. 2005, \apj, 635, 123

\bibitem[{Rao {et~al.}(2003)Rao, Nestor, Turnshek, Lane, Monier, \&
  Bergeron}]{Rao2003}
Rao, S.~M., Nestor, D.~B., Turnshek, D.~A., Lane, W.~M., Monier, E.~M., \&
  Bergeron, J. 2003, \apj, 595, 94

\bibitem[{Rao {et~al.}(2005)Rao, Turnshek, \& Nestor}]{Rao2005}
Rao, S.~M., Turnshek, D.~A., \& Nestor, D.~B. 2005, astro-ph/0509469

\bibitem[{Rownd \& Young(1999)}]{Rownd1999}
Rownd, B.~K., \& Young, J.~S. 1999, \aj, 118, 670

\bibitem[{Ryan-Weber {et~al.}(2003)Ryan-Weber, Webster, \&
  Staveley-Smith}]{Ryan-Weber2003}
Ryan-Weber, E.~V., Webster, R.~L., \& Staveley-Smith, L. 2003, \mnras, 343,
  1195

\bibitem[{Ryan-Weber {et~al.}(2005)Ryan-Weber, Webster, \&
  Staveley-Smith}]{Ryan-Weber2005a}
---. 2005, \mnras, 356, 1600

\bibitem[{Schaye(2001)}]{Schaye2001b}
Schaye, J. 2001, \apjl, 562, L95

\bibitem[{{Solomon} \& {Sage}(1988)}]{Solomon1988}
{Solomon}, P.~M., \& {Sage}, L.~J. 1988, \apj, 334, 613

\bibitem[{{Steidel} {et~al.}(1999){Steidel}, {Adelberger}, {Giavalisco},
  {Dickinson}, \& {Pettini}}]{Steidel1999}
{Steidel}, C.~C., {Adelberger}, K.~L., {Giavalisco}, M., {Dickinson}, M., \&
  {Pettini}, M. 1999, \apj, 519, 1

\bibitem[{Wiklind \& Combes(1999)}]{Wiklind1999}
Wiklind, T., \& Combes, F. 1999, in ASP Conf. Ser. 156: Highly Redshifted Radio
  Lines, Vol. 156, 202

\bibitem[{{Wolfe} {et~al.}(2003){Wolfe}, {Gawiser}, \& {Prochaska}}]{Wolfe2003}
{Wolfe}, A.~M., {Gawiser}, E., \& {Prochaska}, J.~X. 2003, \apj, 593, 235

\bibitem[{Wong \& Blitz(2002)}]{Wong2002}
Wong, T., \& Blitz, L. 2002, \apj, 569, 157

\bibitem[{{Young} \& {Knezek}(1989)}]{Young1989}
{Young}, J.~S., \& {Knezek}, P.~M. 1989, \apjl, 347, L55

\bibitem[{{Young}(2002)}]{Young2002}
{Young}, L.~M. 2002, \aj, 124, 788

\bibitem[{Zwaan {et~al.}(2005{\natexlab{a}})Zwaan, Meyer, Staveley-Smith, \&
  Webster}]{Zwaan2005a}
Zwaan, M.~A., Meyer, M.~J., Staveley-Smith, L., \& Webster, R.~L.
  2005{\natexlab{a}}, \mnras, 359, L30

\bibitem[{Zwaan {et~al.}(2005{\natexlab{b}})Zwaan, van~der Hulst, Briggs,
  Verheijen, \& Ryan-Weber}]{Zwaan2005b}
Zwaan, M.~A., van~der Hulst, J.~M., Briggs, F.~H., Verheijen, M.~A.~W., \&
  Ryan-Weber, E.~V. 2005{\natexlab{b}}, \mnras, 364, 1467

\bibitem[{{Zwaan} {et~al.}(2003)}]{Zwaan2003}
{Zwaan}, M.~A., {et~al.} 2003, \aj, 125, 2842

\end{thebibliography}



\clearpage 

\clearpage

\clearpage

\end{document}